\documentclass{IEEEtran}
\usepackage{cite}
\usepackage{amsmath,amssymb,amsfonts}
\usepackage{algorithmic}
\usepackage{graphicx}
\usepackage{textcomp}
\def\BibTeX{{\rm B\kern-.05em{\sc i\kern-.025em b}\kern-.08em
    T\kern-.1667em\lower.7ex\hbox{E}\kern-.125emX}}
    
\DeclareMathOperator{\sinc}{sinc}

\begin{document}
\title{Inverse Design of Multi-input Multi-output 2D Metastructured Devices}
\author{Luke Szymanski, \IEEEmembership{Student Member, IEEE}, Gurkan Gok, and Anthony Grbic, \IEEEmembership{Fellow, IEEE}
\thanks{ This work has been submitted to the IEEE for possible publication. Copyright may be transferred without notice, after which this version may no longer be accessible. This work was supported by the National Science Foundation, under the Grant Opportunities for Academic Liaison with Industry (GOALI) program under Grant 1807940. (Corresponding author: Anthony Grbic.) }
\thanks{L. Szymanski and A. Grbic are with the Radiation Laboratory, Department of Electrical Engineering and Computer Science at the University of Michigan, Ann Arbor, MI 48109-2122 USA (e-mail: ljszym@umich.edu, agrbic@umich.edu). }
\thanks{G. Gok, is with Raytheon Technologies Research Center, East Hartford, CT 06108 USA (e-mail: Gurkan.Gok@rtx.com).}
}

\maketitle

\begin{abstract}
In this work, an optimization-based inverse design method is provided for multi-input multi-output (MIMO) metastructured devices. Typically, optimization-based methods use a full-wave solver in conjunction with an optimization routine to design devices. Due to the computational cost this approach is not practical for designing electrically-large aperiodic metastructured devices. To address this issue, a 2-D circuit network solver using reduced order models of the metastructure's unit cells is introduced. The circuit network solver is used in conjunction with a gradient-based optimization routine that uses the adjoint variable method to solve large-scale optimization problems like those posed by metastructured devices. To validate the inverse design method, a planar beamformer and an analog signal processor for aperture field reconstruction are designed and validated with full-wave simulations.
\end{abstract}

\begin{IEEEkeywords}
Metastructures, optimization methods, antenna radiation pattern synthesis, analog processing circuits, microwave circuits
\end{IEEEkeywords}

\section{Introduction}
\label{sec:introduction}
\IEEEPARstart{T}{he} need for electromagnetic devices that can perform multiple field transformations, or exhibit multi-input multi-output (MIMO) functionality, arises in many applications such as in antenna beamforming, mode conversion, and recently for analog signal processing \cite{Ozcan, Engheta, Backer}. A promising route to the realization of these devices is through metastructured, or subwavelength textured, devices. Metastructured devices provide large degrees of freedom allowing for a single device to perform multiple field transformations. However, the design of MIMO metastructured devices requires the solution of an inverse design problem. This entails the determination of a set of unknown device characteristics, such as its geometry or material parameters, from a set of known inputs and outputs. Inverse design problems like these often lack direct solution methods and require heuristic or optimization-based methods to be solved. Heuristic methods can significantly simplify the design problem making it analytically tractable. However, they often impose limitations on the possible inputs and outputs, and have inherent errors associated with them. To avoid the limitations of heuristic methods an optimization-based approach is adopted in this work to design MIMO metastructured devices.\\
\indent Previous work in the design of MIMO metastructured devices has included optimization-based design procedures that have been used to realize beamforming networks \cite{Tierney,Sanford}, and analog signal processors, \cite{Engheta,Backer}. The design procedures in \cite{Tierney,Sanford,Engheta} use full-wave solutions to Maxwell's equations to solve the forward problem at every step of the optimization routine. This imposes a high computational cost on these design methods and places practical limitations on their ability to produce large, complex device's. In \cite{Backer}, a method to reduce the computational cost of the forward problem was introduced that used a combination of full-wave solutions and the paraxial approximation to design a cascade of metasurfaces with MIMO functionality. However, only amplitude control of the transmitted field profiles was considered. The aim of this work is to provide an optimization-based procedure for designing MIMO metastructured devices with control over both the amplitude and phase of the transmitted fields.\\
\indent Optimization-based inverse design of metastructured devices presents a computational challenge for two main reasons: (1) The large number of forward problems that need to be solved with different design parameters. (2) The multi-scale nature of the forward problem: subwavelength features in the unit cells and a multi-wavelength device size. The first difficulty is unavoidable so the forward problem solver must be fast. The second difficulty makes the forward problem solver slower, particularly when full-wave solutions are used. Here, these issues are addressed using a fast 2-D circuit network solver that uses reduced-order models for the unit cells of the metastructured device. \\
\indent The design procedure implemented in this work poses the input-output relationship of a MIMO metastructured device as an optimization problem over the design variables. To realize the devices, the optimization problem is solved using a fast 2-D circuit network solver in conjunction with a gradient-based optimization routine that uses the adjoint variable method to efficiently calculate the gradient \cite{Backer,Yablonovitch, Johnson}. The utility of the proposed design procedure is demonstrated through the design of a planar beamformer and an analog signal processor for aperture field decomposition. 
\section{2-D Circuit Network Solver}
In this section, a frequency domain solver for 2-D structures supporting either transverse electric or transverse magnetic polarized fields is introduced. Numerical solutions to these types of problems typically use full-wave methods like finite-difference frequency-domain, finite element methods, or the method of moments. These solution methods are very accurate, however, for electrically-large aperiodic metastructures they are computationally expensive and are impractical to use in their inverse design. Since full-wave solutions are restrictive, alternative solution methods are required to reduce the computational cost of solving the forward problem. Here, this is done by approximating the full-wave solution using reduced-order models of the unit cells and modeling the unit cell interactions with circuit theory. Solving for the device response this way is particularly useful when dealing with port-fed problems like guided-wave structures. In these scenarios, the unit cells can be characterized in isolation to produce good reduced-order models, and the coupling between the unit cells is primarily through their ports. The combination of these attributes allows for the response of complex large structures to be predicted with good accuracy.\\
\begin{figure}[!t]
\centerline{\includegraphics[width=\columnwidth]{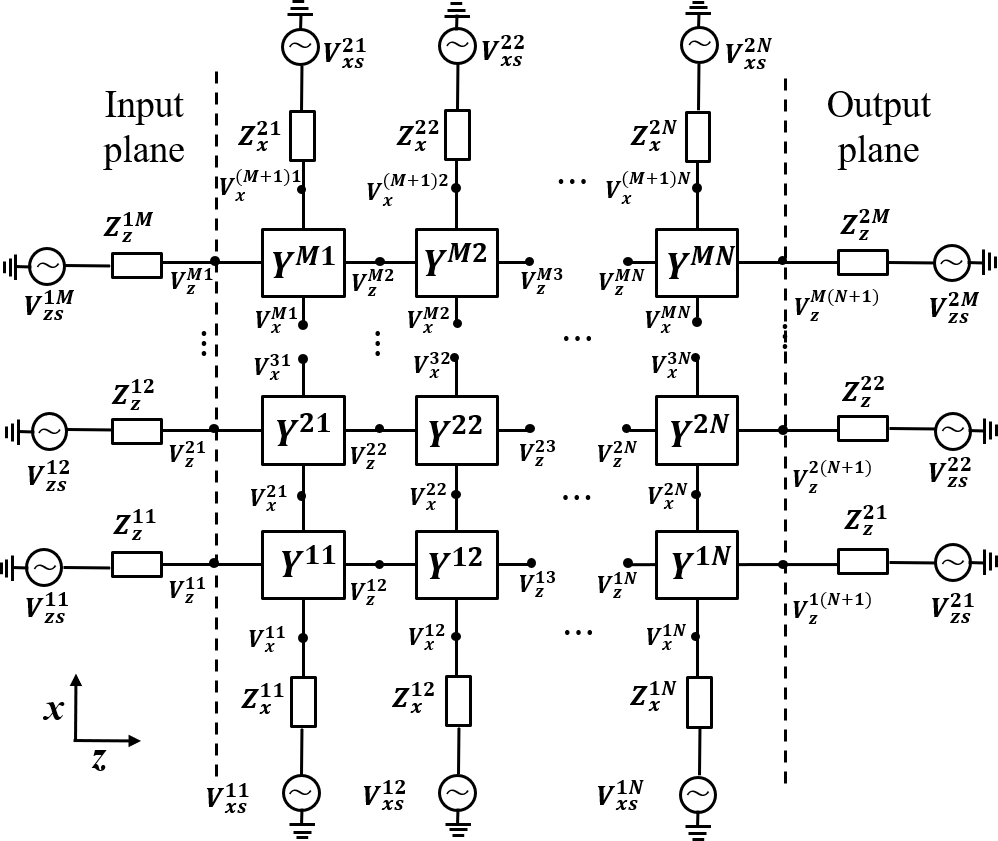}}
\caption{A metastructure (computational domain) consisting of an MxN grid of four-port admittance matrices. The boundary conditions are imposed using lumped element impedances and voltage sources. The voltages at all nodes are solved for by enforcing current conservation (KCL) at each node in the network.}
\label{Ymatrix_network}
\end{figure}
\indent To derive the system of equations that govern the 2-D circuit network solver a unit cell is selected and the device (computational domain) is discretized. Since circuit theory is used to model the unit cell coupling a natural representation for a 2-D unit cell is a general four port network. These four port networks can be represented by scattering matrices, wave matrices, or impedance/admittance matrices, but admittance matrices are used here. Discretizing the computational domain into an MxN grid with unit cells defined by four-port admittance matrices results in the overall circuit shown in Fig. \ref{Ymatrix_network}. It produces a staggered grid of nodal voltages that is organized into two sub-grids: the $V_x$ and $V_z$ grids. The $V_x$ grid represents propagation along the x-direction, and these nodal voltages are referred to as $V^{ij}_x$. The $V_z$ grid represents propagation along the z-direction, and these nodal voltages are referred to as $V^{ij}_z$. Since the computational domain is finite in size, the periphery of the grid is truncated using voltage sources connected in series with lumped impedances. They are used to excite the device and enforce desired boundary conditions.\\
\indent To solve for the voltages in the network, Kirchoff's Current Law (KCL) is imposed at every node in the network. This produces a sparse linear system whose solution is the voltage at every node in the network.
\begin{equation}\label{forward_problem}
 \textbf{v} = \overline{\overline{Q}}^{-1}\textbf{s}
\end{equation}
In \eqref{forward_problem}, $\overline{\overline{Q}}$ characterizes all of the interactions between the unit cells in the network, $\textbf{v}$ is a vector containing all of the nodal voltages, and $\textbf{s}$ is a vector containing the source terms. To determine the structure of $\overline{\overline{Q}}$ and $\textbf{s}$, a grid with M unit cells in the x-direction and N unit cells in the z-direction is considered. There are six types of nodes that need to be accounted for: interior nodes on the $V_x$ grid, interior nodes on the $V_z$ grid, and the nodes on the four boundaries. The two different types of interior nodes and the boundary nodes along the input plane are shown in Fig. \ref{Nodal_eqns}. The three other boundary nodes can be formed in a manner analogous to Fig. \ref{Nodal_eqns} (c).\\
\begin{figure}[!t]
\centerline{\includegraphics[width=\columnwidth]{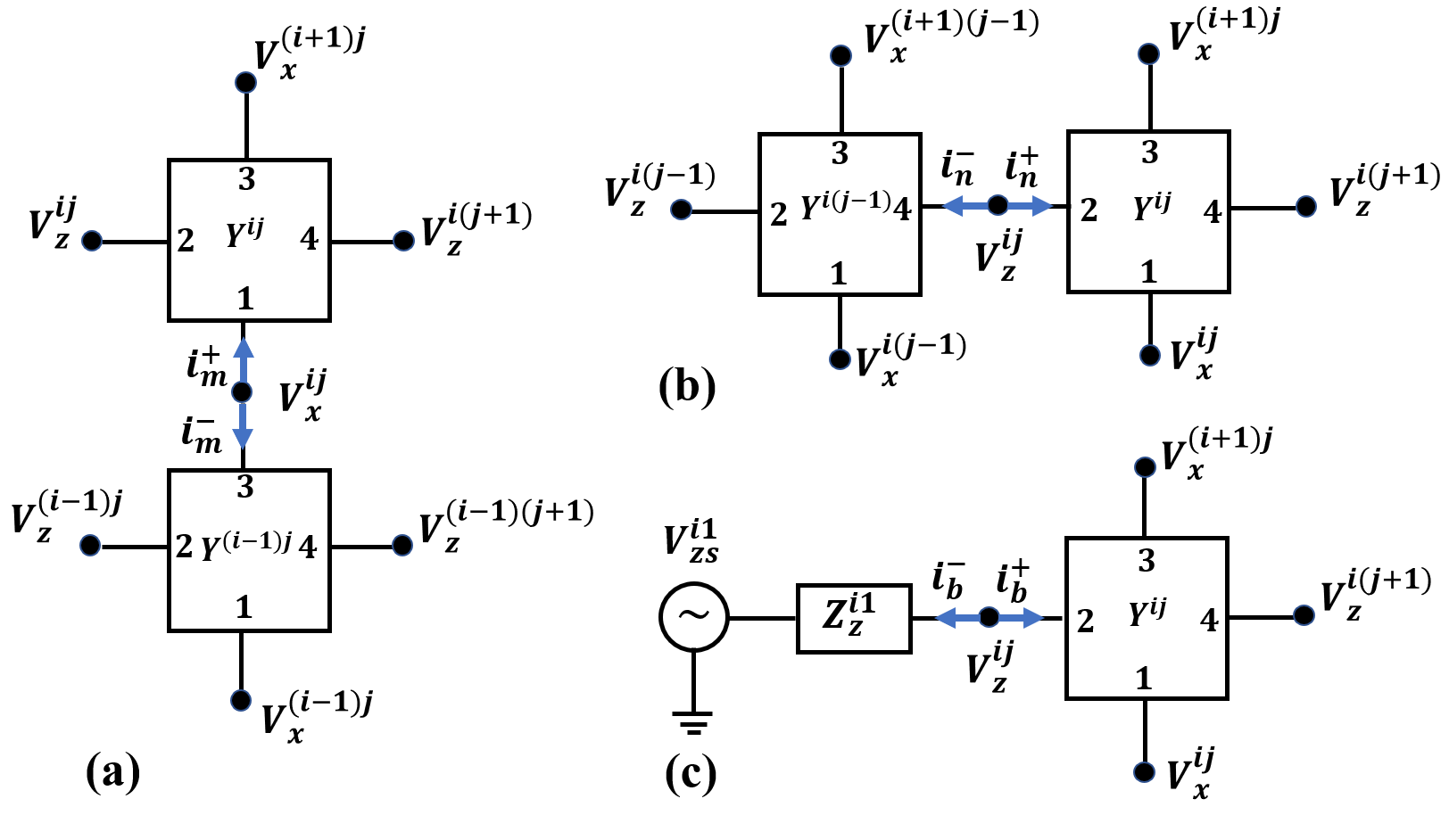}}
\caption{(a) An internal node on the x-grid. (b) An internal node on the z-grid. (c) A boundary node along the input plane in Fig. \ref{Ymatrix_network}. The other boundaries can be obtained in a manner analogous to (c). }
\label{Nodal_eqns}
\end{figure}
\indent The elements of $\overline{\overline{Q}}$ and $\textbf{s}$ in \eqref{forward_problem} are found in the following manner. For a general interior node $V^{ij}_x$ on the $V_x$ grid, shown in Fig. \ref{Nodal_eqns} (a), KCL leads to the following equation,
\begin{equation}\label{Vx_int}
\begin{aligned}
&V_x^{ij}(Y_{11}^{ij}+Y_{33}^{(i-1)j}) + V_x^{(i-1)j}Y_{31}^{(i-1)j} + V_x^{(i+1)j}Y_{13}^{ij}\\
& + V_z^{ij}Y_{12}^{ij} + V_z^{(i-1)j}Y_{32}^{(i-1)j} + V_z^{(i-1)(j+1)}Y_{34}^{(i-1)j} \\
& + V_z^{i(j+1)}Y_{14}^{ij} = 0
\end{aligned}
\end{equation}
Applying KCL at a general interior node $V^{ij}_z$ on the $V_z$ grid, shown in Fig. \ref{Nodal_eqns} (b), leads to the following equation,
\begin{equation}\label{Vz_int}
\begin{aligned}
&V_z^{ij}(Y_{22}^{ij}+Y_{44}^{i(j-1)}) + V_z^{i(j-1)}Y_{42}^{i(j-1)} + V_z^{i(j+1)}Y_{24}^{ij}\\
& + V_x^{ij}Y_{21}^{ij} + V_x^{(i+1)j}Y_{23}^{ij} + V_x^{(i+1)(j-1)}Y_{43}^{i(j-1)} \\
& + V_x^{i(j-1)}Y_{41}^{i(j-1)} = 0
\end{aligned}
\end{equation}
Applying KCL at the four types of boundary nodes (see Fig. \ref{Ymatrix_network}) produces the following equations: the left boundary $V_z^{i1}$ (shown in Fig. \ref{Nodal_eqns} (c)),
\begin{equation}\label{Vzin_Boundary}
V_z^{i1}(Y_{22}^{i1}+\frac{1}{Z_z^{i1}}) + V_z^{i2}Y_{24}^{i1} + V_x^{i1}Y_{21}^{i1} + V_x^{(i+1)1}Y_{23}^{i1} = \frac{V_{zs}^{i1}}{Z_z^{i1}} 
\end{equation}
the right boundary $V_z^{i(N+1)}$,
\begin{equation}\label{Vzout_Boundary}
\begin{aligned}
V_z^{i(N+1)}(Y_{44}^{iN}+\frac{1}{Z_z^{i2}}) + V_z^{iN}&Y_{42}^{iN} + V_x^{iN}Y_{41}^{iN}\\ 
& + V_x^{(i+1)N}Y_{43}^{iN} = \frac{V_{zs}^{i2}}{Z_z^{i2}} 
\end{aligned}
\end{equation}
the bottom boundary $V_x^{1j}$,
\begin{equation}\label{Vxb_Boundary}
\begin{aligned}
V_x^{1j}(Y_{11}^{1j}+\frac{1}{Z_x^{1j}})+V_x^{2j}Y_{13}^{ij} + V_z^{1j}Y_{12}^{1j} + &V_z^{1(j+1)}Y_{14}^{1j}\\
& = \frac{V_{xs}^{1j}}{Z_x^{1j}} 
\end{aligned}
\end{equation}
the top boundary $V_x^{(M+1)j}$,
\begin{equation}\label{Vxt_Boundary}
\begin{aligned}
V_x^{(M+1)j}(Y_{33}^{Mj}+\frac{1}{Z_x^{2j}}) + V_x^{Mj}&Y_{31}^{Mj} + V_z^{Mj}Y_{32}^{Mj}\\
& + V_z^{M(j+1)}Y_{34}^{Mj} = \frac{V_{xs}^{2j}}{Z_x^{2j}}.
\end{aligned} 
\end{equation}
Expressing KCL at every node in the network using \eqref{Vx_int}-\eqref{Vxt_Boundary} forms a linear system of $2MN+M+N$ equations shown in \eqref{forward_problem}. The matrix $\overline{\overline{Q}}$ can be quite large as its dimensions are $(2MN+M+N)\times(2MN+M+N)$. However, there are a maximum of seven non-zero terms in each row of $\overline{\overline{Q}}$. Therefore, when its dimensions are large it is sparse. This sparsity allows for the total device response, $\textbf{v}$, of large aperiodic metastructured devices to be rapidly evaluated once the admittance matrices of its unit cells are characterized. \\
\indent The major advantage of solving for the device response this way is that it can maintain a high-level of accuracy if good models of the unit cells are developed while avoiding full-wave solutions at run time. Eliminating full-wave solutions reduces the computational cost of solving the forward problem significantly, making the circuit network solver useful for the optimization-based inverse design of large aperiodic metastructures. The solver does come with limitations though. One limitation is the requirement that the unit cells can be represented as a four port networks. This means that the problem of interest's unit cells must have an equivalent guided wave representation, which is not always possible if there is a continuous spectrum of propagating waves. Another limiting assumption is that all of the interactions between the unit cells can be captured using a single guided mode. Meaning that mutual coupling between the unit cells and higher-order modes excited by inclusions or discontinuities are neglected. If these interactions become significant this assumption can be relaxed, and multi-modal Y-matrices or wave matrices can be used to capture these effects, \cite{Mesa,Alsolamy}. Using higher-order modes is not always necessary and does increase the computational cost which, is why they were not used here.
\section{MIMO Inverse-design Procedure}
\begin{figure}[!t]
\centerline{\includegraphics[width=\columnwidth]{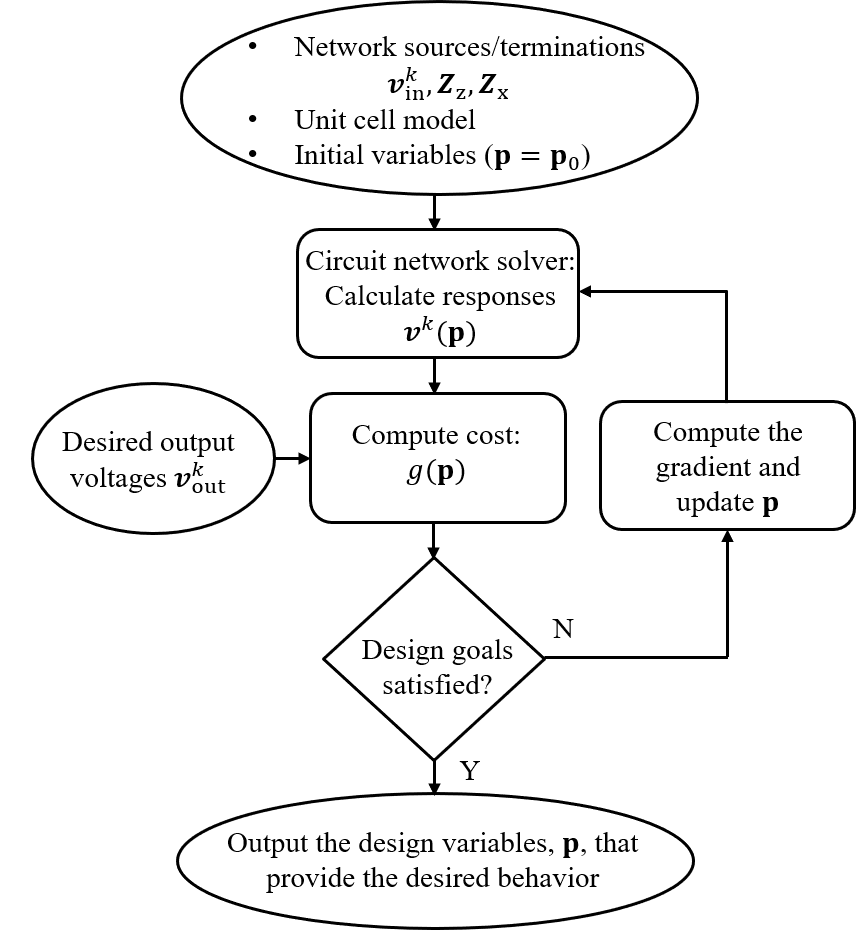}}
\caption{Flowchart depicting the inverse-design procedure. }
\label{Flowchart}
\end{figure}

\indent The fast 2-D circuit network solver introduced in the previous section solves for the output of a device given an input and all of the unit cell's admittance parameters. However, in a MIMO inverse-design problem the unit cell's admittance parameters are solved for given a set of desired inputs and outputs. In this section, an optimization-based inverse design procedure is described that achieves this goal. The desired MIMO functionality is realized by formulating the design objectives as an optimization problem that can be solved using the 2-D circuit network solver in conjunction with an off-the-shelf optimization routine. The design procedure is outlined in Fig. \ref{Flowchart}, and the details of how the optimization problem is formulated and solved is provided in the following subsections.
\subsection{Optimization Problem}
\indent The design of a multi-input multi-output device begins with a set of inputs and outputs that describe it's functionality. Here, these inputs and outputs are voltage distributions along the input and output planes of the device, see Fig. \ref{Ymatrix_network}. These voltage distributions will be referred to as $\{\textbf{v}^k_\textrm{in}\}$ for the inputs and $\{\textbf{v}^k_\textrm{out}\}$ for the outputs where, $k\in\{1,2,...,K\}$ and $K$ is the total number of input-output pairs. Here, the term input-output pair refers to an input voltage distribution and its associated output voltage distribution, $(\textbf{v}^k_\textrm{in},\textbf{v}^k_\textrm{out})$. Specifically, $\textbf{v}^k_\textrm{out}$ are the observed voltages when the network is excited by $\textbf{v}^k_\textrm{in}$.  To realize the MIMO network described by $\{\textbf{v}^k_\textrm{in}\}$ and $\{\textbf{v}^k_\textrm{out}\}$ using optimization, the device's performance for each input-output pair must be represented by a single real number. This can be done with the following cost function for the kth input-output pair,
\begin{equation}
g_k(\textbf{p}) = \frac{1}{2}(\textbf{v}^k(\textbf{p})-\textbf{v}^k_\textrm{out})^H\overline{\overline{G}}(\textbf{v}^k(\textbf{p})-\textbf{v}^k_\textrm{out})
\label{cost_k}
\end{equation}
where $\textbf{p}$ is a vector containing all of the design variables in the network, the vector $\textbf{v}^k(\textbf{p})$ contains the voltages in the network (subject to the design variables) when it is excited by $\textbf{v}_\textrm{in}^k$, and the superscript $H$ indicates the conjugate transpose. The matrix $\overline{\overline{G}}$ is diagonal and positive-semidefinite. It is used to select and scale the elements of $\textbf{v}^k(\textbf{p})-\textbf{v}^k_\textrm{out}$. The performance of the device over all of the input-output pairs is determined by summing \eqref{cost_k} over $k$ to produce the total cost function,
\begin{equation}
g(\textbf{p}) = \sum_{k=1}^K g_k(\textbf{p})
\label{cost_total}
\end{equation}
Framing the problem in this way reduces the design of the multi-input multi-output network to finding the design that minimizes the total cost, i.e. the minimizer $ \textbf{p}^* $, of \eqref{cost_total}. This goal is represented by the following optimization problem,
\begin{equation}\label{Optim_prob}
\begin{aligned}
& \arg \min_\textbf{p} g(\textbf{p})\\ 
&\textrm{subject to}: \ \ \textbf{p}_\textrm{lb} \preceq \textbf{p} \preceq \textbf{p}_\textrm{ub}
\end{aligned}
\end{equation}
where $\textbf{p}_\textrm{lb}$ and $\textbf{p}_\textrm{ub}$ are vectors containing the lower and upper bounds of the design variables, respectively. \\
\indent Now that the optimization problem \eqref{Optim_prob} has been posed an appropriate optimization algorithm needs to be selected to solve it. Since the optimization problem is in general non-convex, local optimization or gradient-based algorithms are not guaranteed to find globally optimal solutions. However, global optimization algorithms do not tend to perform well in high-dimensional design spaces like the design space of a metastructured device. In high-dimensional spaces, local methods tend to outperform global methods when they use gradient information to navigate the design space. However, if the gradient cannot be expressed explicitly in closed form and the dimensions of the design space are large the computational cost of calculating the gradient can become prohibitive. To avoid this issue the adjoint variable method \cite{Backer,Yablonovitch, Johnson} can be used to calculate the gradient at a reduced computational cost. For these reasons, a gradient-based optimization routine utilizing the adjoint variable method is chosen to solve \eqref{Optim_prob}. Since it is a local method there is no guarantee of convergence to a globally optimal solution, i.e. the solution of \eqref{Optim_prob}. However, this is not a problem since the globally optimal design is not required. The required design is just one that meets the design specifications so, a solution to \eqref{Optim_prob} is considered any design that satisfies the design specifications.
\subsection{Adjoint Variable Method}
\indent  As discussed in the previous section, the optimization problem, \eqref{Optim_prob}, will be solved using an off-the-shelf gradient-based optimization routine. Here, MATLAB's constrained optimization routine fmincon() is used. To improve the performance of the algorithm on large-scale problems, a user defined gradient is implemented that is calculated using the adjoint variable method. To motivate the use of the adjoint variable method, a comparison between calculating the gradient using it and using finite-differences is considered. Consider the design of a device with $P$ design variables and $K$ input-output pairs. Calculating the gradient of \eqref{cost_total} using finite-differences requires $(P+1)\times K$ forward solutions to \eqref{forward_problem}. When $P$ becomes large this is not a practical means of determining the gradient especially if the dimensions of $\overline{\overline{Q}}$ are also large. On the other hand, if the adjoint variable method is used only $2\times K$ forward solutions are required. This reduction in the number of forward problem solutions significantly accelerates the optimization routine, and is critical to enabling the inverse design of electrically-large aperiodic metastructures. \\
\indent The implementation of the adjoint variable method presented here is a quasi-analytical technique for calculating the gradient of \eqref{cost_total}. It requires solutions to the forward problem, \eqref{forward_problem} and an adjoint problem associated with \eqref{cost_total}, \cite{Johnson_adjoint}. The adjoint problem associated with \eqref{cost_total} is formed by observing that the gradient of \eqref{cost_k} can be expressed as, 
\begin{equation} \label{grad_cost_k}
\nabla_\textbf{p}(g_k(\textbf{p})) = -\Re\{(\textbf{v}^k-\textbf{v}^k_\textrm{out})^H\overline{\overline{G}}\ \ \overline{\overline{Q}}_k^{-1} \overline{\overline{V}}^k_\textbf{p}  \}
\end{equation}
where $\overline{\overline{V}}^k_\textbf{p}$ is the following matrix,
\begin{equation}
\overline{\overline{V}}^k_\textbf{p} = (\frac{\partial \overline{\overline{Q}}_k}{\partial p_1}\textbf{v}^k | \frac{\partial \overline{\overline{Q}}_k}{\partial p_2}\textbf{v}^k | \frac{\partial \overline{\overline{Q}}_k}{\partial p_3}\textbf{v}^k | \ldots |\frac{\partial \overline{\overline{Q}}_k}{\partial p_P}\textbf{v}^k)
\end{equation}
This matrix can be solved for analytically if expressions for the derivatives of the admittance matrix (Y-matrix) elements in $\overline{\overline{Q}}_k$, the $\overline{\overline{Q}}$ matrix for the kth input-output pair, are available to determine $\frac{\partial \overline{\overline{Q}}_k}{\partial p_i}$. Otherwise, it can be obtained using finite-differences to approximate $\frac{\partial \overline{\overline{Q}}_k}{\partial p_i}$ at a low computational cost. The efficiency of calculating the gradient using \eqref{grad_cost_k} can be improved by observing that the product on the right hand side of \eqref{grad_cost_k}, excluding $\overline{\overline{V}}^k_\textbf{p}$, forms a vector, $\boldsymbol{\lambda}_k^H$, that can be solved for independently,
\begin{equation} \label{adjoint_variable}
\boldsymbol{\lambda}_k^H = (\textbf{v}^k-\textbf{v}^k_\textrm{out})^H\overline{\overline{G}}\ \ \overline{\overline{Q}}_k^{-1}
\end{equation}
Rearranging this expression yields the adjoint problem associated with \eqref{cost_k},
\begin{equation}
\overline{\overline{Q}}_k^{H}\boldsymbol{\lambda_k} = \overline{\overline{G}}(\textbf{v}^k-\textbf{v}^k_\textrm{out}).
\end{equation}
This allows for the adjoint variable $\lambda_k$ to be computed at the cost of solving a forward problem of equal complexity to the original problem \eqref{forward_problem}. Using \eqref{adjoint_variable} in \eqref{grad_cost_k} yields,
\begin{equation}\label{gradient_adjoint_k}
\nabla_\textbf{p}(g_k(\textbf{p})) = -\Re\{\boldsymbol{\lambda}_k^H \overline{\overline{V}}^k_\textbf{p}  \}
\end{equation}
Expressing \eqref{grad_cost_k} in this way provides a means of obtaining the gradient of the kth cost function, \eqref{cost_k}, at the computational expense of effectively two forward problem solutions. The gradient of \eqref{cost_total} is then obtained by summing \eqref{gradient_adjoint_k} over $k$,
\begin{equation}\label{Total_gradient}
\nabla_\textbf{p}(g(\textbf{p})) = -\sum^K_{k=1} \Re\{\boldsymbol{\lambda}_k^H \overline{\overline{V}}^k_\textbf{p}  \}.
\end{equation}
Therefore, in a problem containing $P$ design variables the gradient can be determined with $2\times K$ forward problem solutions using \eqref{Total_gradient}. Rather than the $(P+1)\times K$ forward problem solutions required by finite-differences.

\begin{figure}[!t]
\centerline{\includegraphics[width=\columnwidth]{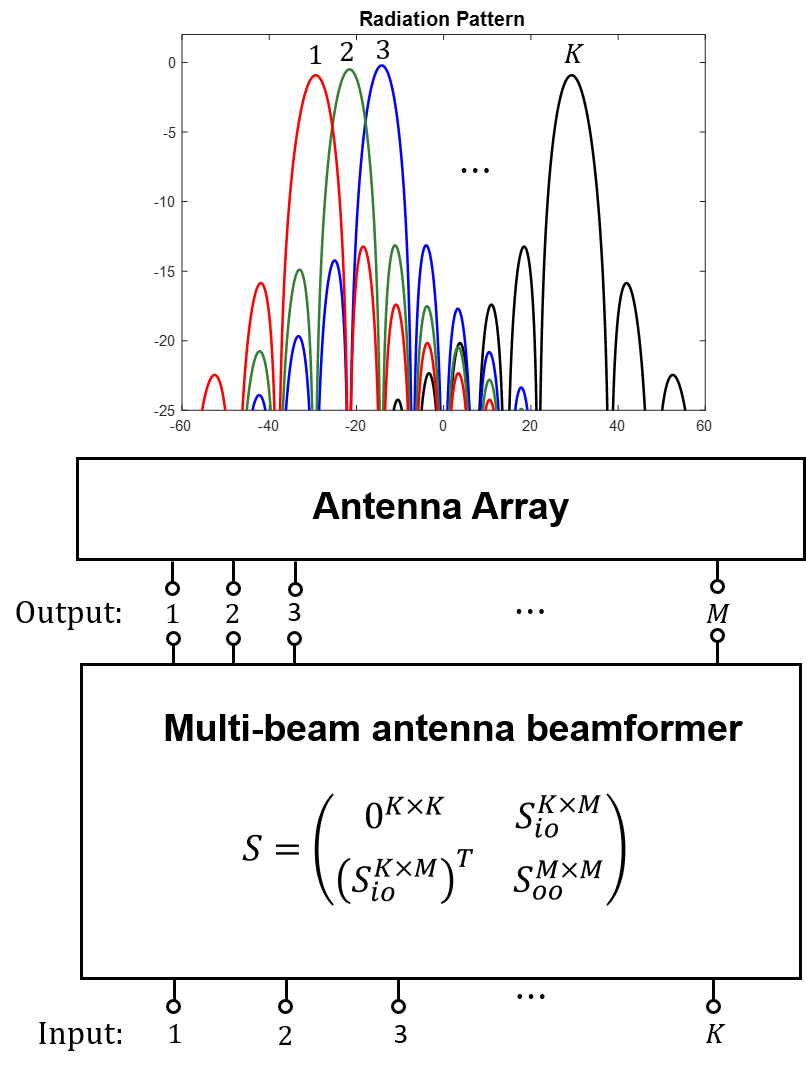}}
\caption{A block diagram of a multi-beam antenna system. It is composed of an antenna array and a reciprocal beamformer that can produce $K$ beams simultaneously. The upper left block of the beamformer's S-matrix is $0^{K\textrm{x}K}$ indicating that its input ports are impedance matched and decoupled. The block $S_{io}^{K\textrm{x}M}$ determines the aperture field produced by exciting each of the input ports. The elements of $S_{oo}^{M\textrm{x}M}$ are free variables, and are neglected in the design of the beamformer.}
\label{Multibeam_antenna}
\end{figure}
\section{Planar Metastructured Beamformer}
\indent To demonstrate the effectiveness of the design procedure presented in Section III the multi-beam antenna beamformer shown in Fig. \ref{Multibeam_antenna} is designed. The beamformer designed using this framework provides several advantages over Butler matrices and quasi-optical beamformers such as the Rotman lens, \cite{Rotman}, the Luneburg lens, \cite{Luneburg}, or beamformers based on transformation optics \cite{Gok}. The advantage of the proposed design procedure is that it allows for perfect control of both the amplitude and phase of every aperture field. In contrast, a Rotman lens lacks amplitude control and provides perfect phasing for a maximum of three aperture fields. The Luneburg lens produces identical aperture fields for all scan angles but the amplitude pattern cannot be controlled. To control amplitude, the design method reported in \cite{Gok} can be used. However, this design method allows for perfect phasing and amplitude control of only one aperture field. Butler matrices provide an alternative to quasi-optical beamformers, and they can theoretically produce an arbitrary number of perfectly phased uniform amplitude aperture fields, \cite{Butler}. However, Butler matrices are limited to uniform aperture fields for single port excitations. Whereas beamformers designed using the procedure presented in Section III have no inherent restrictions on the possible aperture fields, and could be advantageous when designing beamformers for antenna arrays with deeply subwavength element spacings and high input impedances like tightly coupled or connected arrays, \cite{Hansen}.
\subsection{Design Specifications}
\indent The beamforming region is assumed to be lossless and reciprocal and is designed to produce nine beams that are each associated with a different input port. It will operate at 10GHz and is intended for use with an aperture antenna that has a width of $W_{ap}=8\lambda_0$. To produce a close approximation to a continuous aperture field the spacing between the output ports is chosen to be $\lambda_0/10$. This determines the discretization of the beamforming region and thus the unit cell size. The width of the beamformer is the same as the width of the aperture antenna, $8\lambda_0$, and the depth is chosen to be $2.4\lambda_0$. This depth was selected to provide sufficient distance to spread out the input power without utilizing cavity effects from the edges of the beamformer. Therefore, the overall dimensions of the beamformer are $8\lambda_0\times2.4\lambda_0$. This corresponds to a network with $M=80$ unit cells in the transverse direction and $N=24$ in the longitudinal direction.\\
\indent The nine inputs to the network are $70\Omega$ port excitations located along the input plane, shown in Fig. \ref{Ymatrix_network}. The inputs are spaced by $0.8\lambda_0$ starting from the center line of the beamformer. This spacing aids in isolating the input ports from each other: a requirement for the simultaneous excitation of the beams. The requirement of isolation between the input ports and losslessness mandate that the radiated fields are orthogonal over a period of the radiation pattern \cite{Allen,White,Stein}. This restricts the possible aperture fields and must be considered when selecting the desired radiation patterns of the aperture antenna. A well known set of functions that satisfy mutual orthogonality are sinc functions with appropriate angular spacings. For this reason, the aperture fields are chosen to have uniform amplitude with linear phase gradients corresponding to the following tangential wavenumbers,
\begin{equation}
k_n = \frac{2\pi n}{Md}, \ n\in\{0,\pm 1,\pm 2,...,\pm (M-1)\}.
\label{orthogonality_relationship}
\end{equation}
In \eqref{orthogonality_relationship}, $d$ is the spacing between the output ports and $M$ is the total number of output ports. In this design the nine beams correspond to $n=0,\pm 1,\pm 2,\pm 3, \pm 4$. Since $d=\lambda_0/10$ and $M=80$, these correspond to beams at $\theta_n = 0^\circ, \pm 7.18^\circ, \pm 14.48^\circ, \pm 22.02^\circ, \pm 30^\circ$. The output terminations are given by the input impedance of the aperture antennas ports. The input impedance for each port is $140\Omega$ at broadside so, the output terminations for each of the $n$ excitations are given by $140\Omega/\cos{\theta_n}$. The remaining ports in Fig. \ref{Ymatrix_network} along the top and bottom of the beamformer, as well as the unused ports along the input plane are terminated in open circuits. Alternatively, the exact termination presented by each antenna at the output ports could be included directly in the design process. This would only require a slight modification to the forward problem solver. An $M$-port termination at the output plane would need to be included altering \eqref{Vzout_Boundary}. Then an $M$-port admittance matrix of the antenna could be characterized and included in the design process. 
\begin{figure}[!t]
\centerline{\includegraphics[width=\columnwidth]{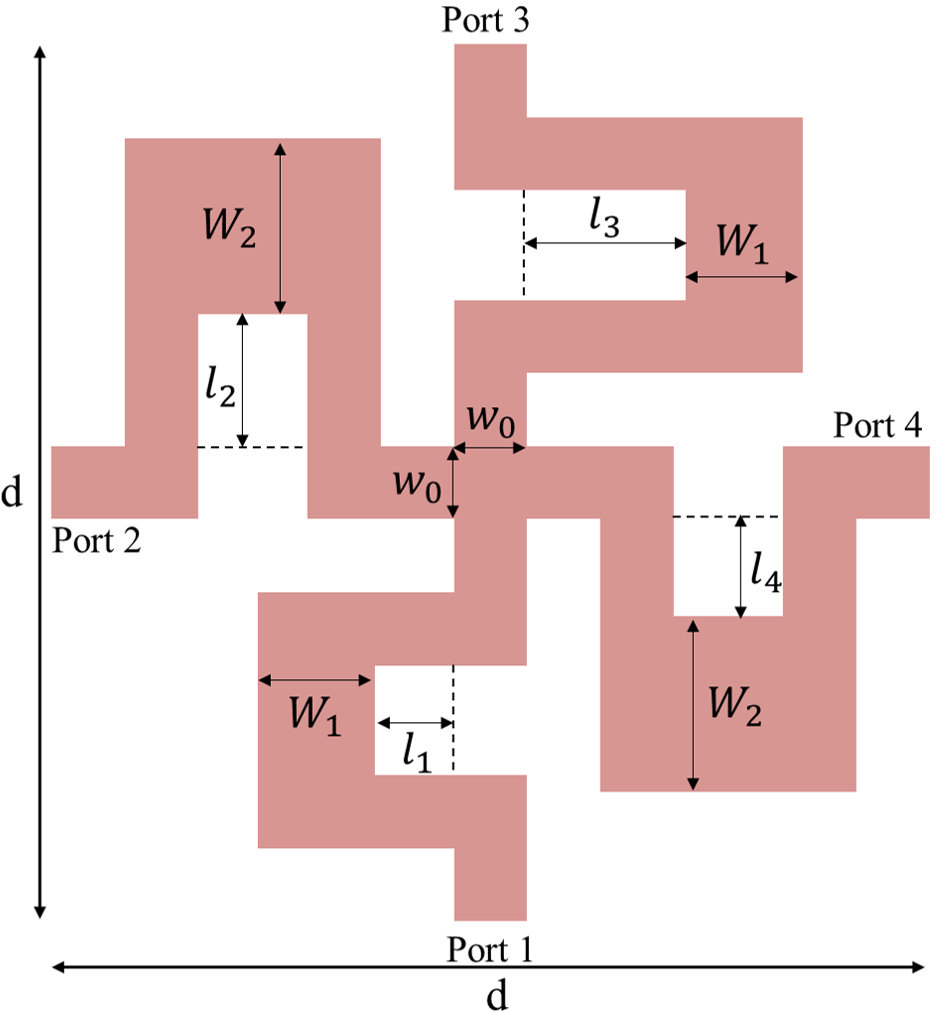}}
\caption{The microstrip unit cell used in the design of the metastructured beamformer and the analog signal processor. The square unit cell has dimensions $d$x$d$. It possesses six degrees of freedom, $W_1,\ W_2,\ l_1,\ l_2,\ l_3,\ \textrm{and} \ l_4$ that can be varied to realize a variety of four-port admittance matrices. The microstrip lines at the input ports, central junction, and the lines corresponding to $l_1,\ l_2,\ l_3,\  \textrm{and}\ l_4$ have a constant width of $w_0$.}
\label{Unit_cell} 
\end{figure}
\begin{figure*}[!t]
\centerline{\includegraphics[width=\textwidth]{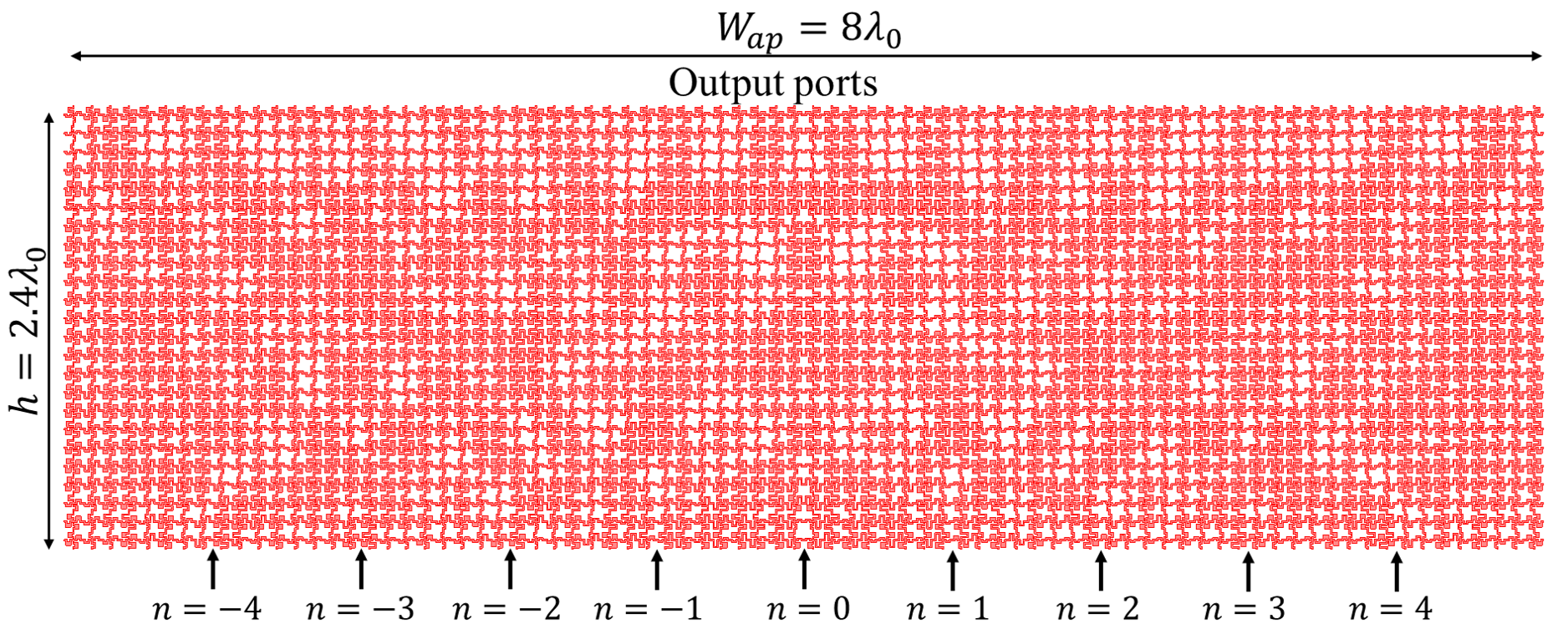}}
\caption{The patterned metastructured beamformer produced by the proposed inverse-design procedure. There are nine input ports that each produce unique voltage distributions across the 80 output ports to form the desired aperture fields. These aperture fields produce beams at $\theta_n = 0^\circ, \pm 7.18^\circ, \pm 14.48^\circ, \pm 22.02^\circ, \pm 30^\circ$. The beamformer is designed to work at 10GHz and  is composed of 1920 unit cells. The width of the aperture is $W_{ap} = 8\lambda_0 \ (24\textrm{cm})$ and the beamformer has a depth of $h = 2.4\lambda_0 \  (7.2\textrm{cm})$. }
\label{MSTL_beamformer}
\end{figure*}
\begin{figure}[!t]
\centerline{\includegraphics[width=\columnwidth]{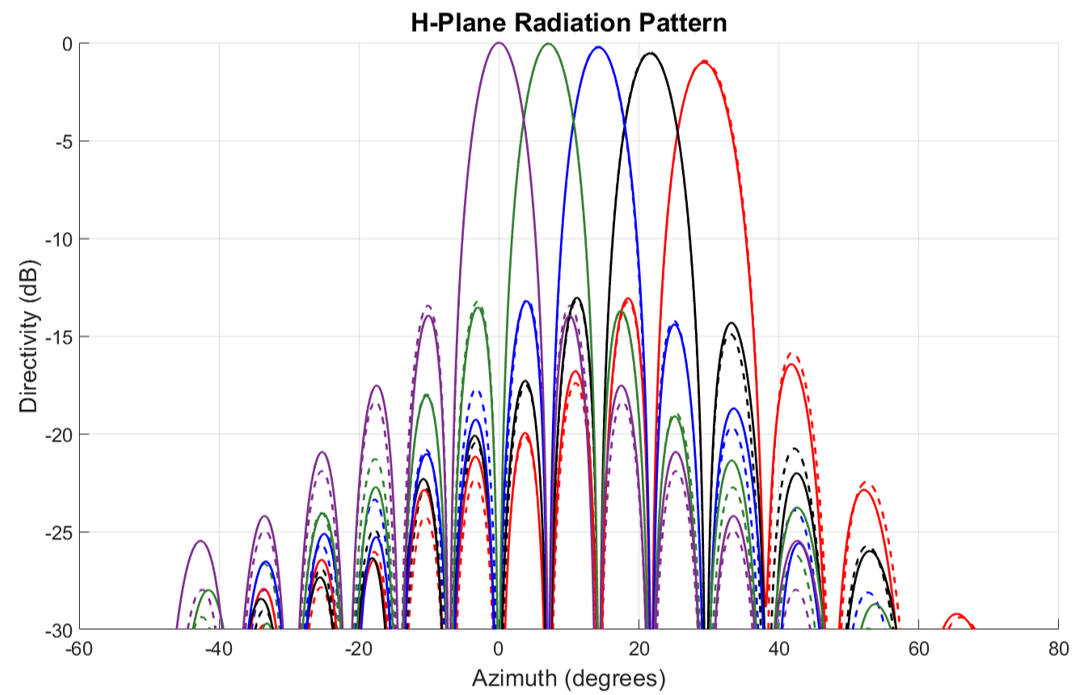}}
\caption{Radiation from an aperture antenna excited by the output voltages of the beamforming network: dashed lines correspond to the radiation pattern from the desired voltages and the solid lines correspond to the voltages calculated using the circuit network solver. The radiation pattern is computed using MATLAB by assuming a piece-wise uniform electric field across the aperture. The total radiated electric field is calculated using equivalent magnetic currents across the aperture. For clarity, only beams corresponding to positive scan angles are shown since the beams corresponding to negative scan angles are identical due to symmetry.}
\label{9_beam_radPat_Y}
\end{figure}
\begin{figure}[!t]
\centerline{\includegraphics[width=\columnwidth]{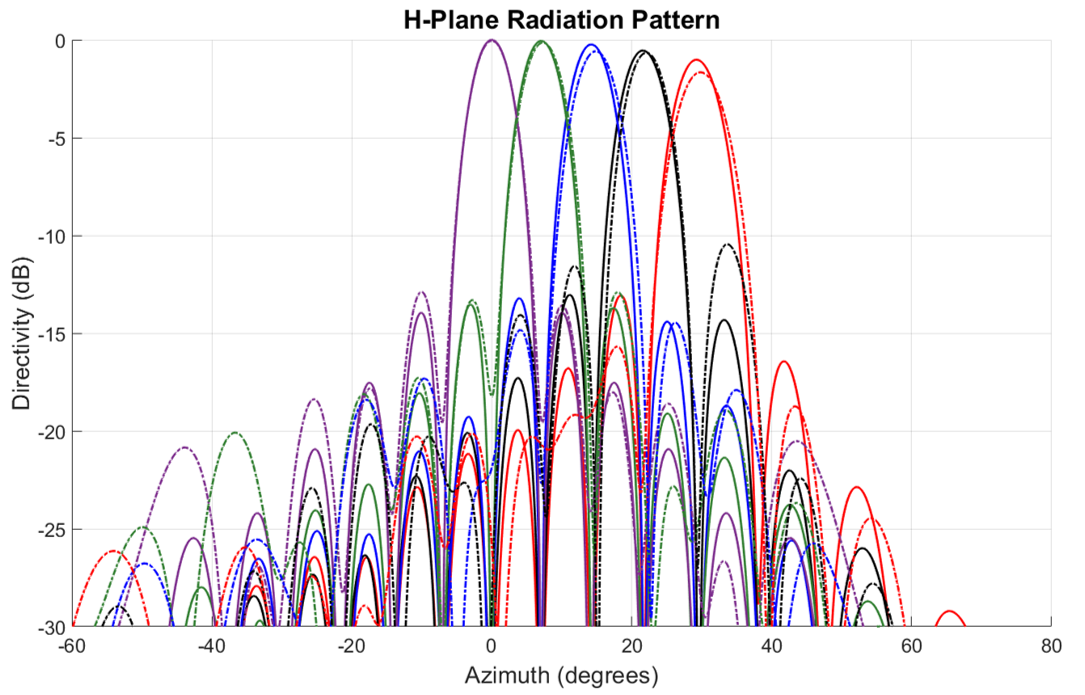}}
\caption{Radiation from an aperture antenna excited by the output voltages of the beamforming network: solid lines correspond to the radiation pattern from the voltages calculated using the circuit network solver and the dot-dashed lines correspond to the voltages calculated using the full-wave results. The radiation pattern is computed using MATLAB by assuming a piece-wise uniform electric field across the aperture. The total radiated electric field is calculated using equivalent magnetic currents across the aperture. For clarity, only beams corresponding to positive scan angles are shown since the beams corresponding to negative scan angles are identical due to symmetry.}
\label{9_beam_radPat_FW}
\end{figure}
\subsection{Unit Cell Design}
\indent In order to use the design procedure presented in Section III, a suitable unit cell must be chosen and characterized. It is desirable for the unit cell to contain the maximum number of degrees of freedom (design variables) to allow for extreme field transformations with a compact design. An arbitrary lossless and reciprocal four-port admittance matrix has a maximum of ten degrees of freedom. However, due to field averaging arguments, the degrees of freedom are reduced to effective material parameters in electrically-small structures,\cite{Pestourie,Holloway}. The beamformer can be viewed as a lossless, reciprocal, polarization conserving medium supporting a TE wave in the x-z plane. The most general medium supporting this type of propagation is a 2-D omega medium that conserves polarization, i.e. a medium with the following material properties,
\begin{align*}
\overline{\overline{\mu}} = 
\begin{pmatrix}
\mu_{xx} & \mu_{xz}\\
\mu_{xz} & \mu_{zz}
\end{pmatrix}
, \ \
\varepsilon_{yy}
, \ \
\overline{\overline{a}} = \overline{\overline{b}} = j 
\begin{pmatrix}
0 & a_{xy} & 0 \\
-a_{xy} & 0 & a_{zy}\\
0 & -a_{zy} & 0
\end{pmatrix}
\end{align*}
where, $\overline{\overline{a}}$ and $\overline{\overline{b}}$ are the magneto-electric and electro-magnetic tensors, respectively. This perspective provides some guidance for designing the unit cell. It points to the fact that a maximum of six design variables should be included in the unit cell. Additional design variables will only complicate the characterization of the unit cell without producing observable degrees of freedom in the response. It also provides information on what characteristics of the unit cell the design variables should affect. The design variables should control: (1) The path lengths in the x and z directions, as well as coupling between the two directions to change $\overline{\overline{\mu}}$. (2) The transmission-line widths to change $\varepsilon_{yy}$. (3) The asymmetry in the x and z directions changes the bianisotropic response, $\overline{\overline{a}}$.\\
\indent For these reasons the unit cell depicted in Fig. \ref{Unit_cell} is selected. The unit cell has dimensions $d\textrm{x}d$ where $d = 3\textrm{mm}$. It is composed of four branches of microstrip transmission-lines meeting at a cross junction in the center. To avoid parasitic effects when interconnecting the unit cells, the transmission-line width at all of the ports is a constant value of $w_0 = 0.25\textrm{mm}$. There are six degrees of freedom in the unit cell: four lengths and two widths. Each branch contains one of the four length variables. While one of the width variables is in the x-directed branches and the other width variable is in the the z-directed branches.\\
\indent To design the beamformer, the unit cell admittance parameters need to be available as continuously differentiable functions of the lengths and widths. Here, this is achieved by constructing a reduced-order model of the unit cell from a database of full-wave simulations. The unit cells in the database were characterized in isolation using Keysight's Momentum, and a model of Rogers RO5880 substrate neglecting dissipative losses was used. The substrate has a height of $h = 0.787\textrm{mm}$ and a relative permittivity $\varepsilon_r = 2.2$.  The design variables for each unit cell in the database are chosen to form a uniformly spaced grid of the allowable lengths and widths in the design space. The allowable range of widths is $0.2\leq W_{1,2}\leq 0.8\textrm{mm}$ and the lengths vary between $0\leq l_i \leq l^{max}_i$. Where $l^{max}_i = d/2-w_0/2-W$ and $W$ corresponds to the width of the line which is connected to $l_i$ for $i\in {\{1,2,3,4}\}$. The admittance parameters from the database are then spline interpolated to generate a model of the unit cell that is differentiable with respect to the variable lengths and widths.
\subsection{Optimization and Results}
\indent Since the beamformer produces symmetric beams, the beamformer should be symmetric as well. For this reason symmetry is imposed across the center line in the forward problem solver. This results in the device having 5,760 design variables. The design variables are solved for by providing the optimization routine with the input and output terminations, the desired input and output voltage distributions, the model of the unit cell, and an initial guess of uniform lengths and widths for the design variables. The algorithm then searches the design space to find a set of design variables that satisfies the design requirements. This process took $\sim6$ hours to produce a satisfactory design on a personal computer with an i7-9700 CPU @ 3GHz w/8 cores and 64GB RAM. The layout of the design is shown in Fig. \ref{MSTL_beamformer}.\\
\indent To evaluate the performance of the beamformer, the radiation pattern of the multi-beam antenna is computed using MATLAB. The calculation assumes that the electric field across the aperture has a piece-wise uniform amplitude and phase. The equivalent magnetic currents are then determined and the total radiated electric field is calculated. The radiation patterns resulting from the circuit network solver (forward problem solver) are in excellent agreement with the desired beams, as shown in Fig. \ref{9_beam_radPat_Y}. The beamformer inputs have a worst case isolation of $21.2\textrm{dB}$ and a worst case reflectance of $-38\textrm{dB}$ for the broadside excitation. These results are then compared to a full-wave simulation of the beamformer performed in Keysight Momentum using the same substrate that was used in the characterization of the unit cell. The simulation, the equivalent of one forward problem solution using Momentum, took approximately 98 hours on a high-performance computing cluster with access to 15 cores and 600 GB RAM. The results of the simulation are shown in Fig. \ref{9_beam_radPat_FW}. Due to slight variations in the observed voltages at the beamformer ports, there are small shifts and a slight broadening of the beams. The isolation and input impedance match are slightly degraded as well. There is a worst case isolation of $16.2\textrm{dB}$ and a maximum reflectance of $-15.2\textrm{dB}$ for the broadside excitation. However, the overall agreement is quite good between the full-wave and circuit network solver results.
\begin{figure}[!t]
\centerline{\includegraphics[width=\columnwidth]{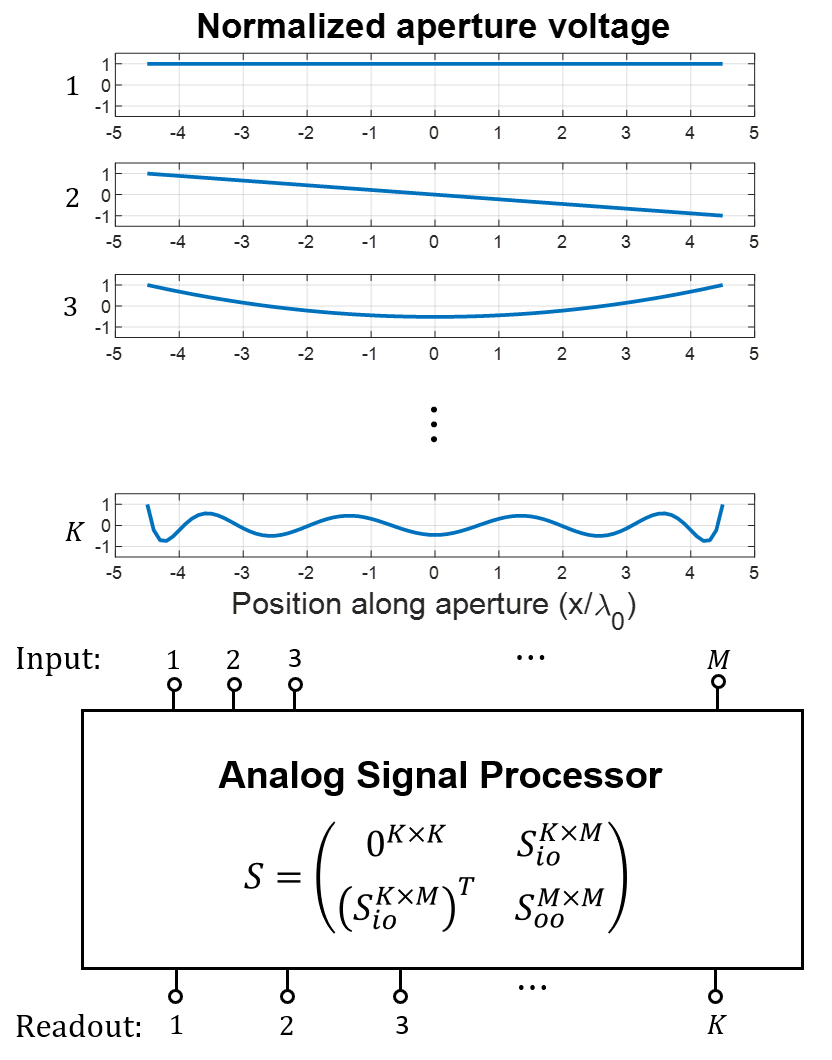}}
\caption{A depiction of the analog signal processor that performs aperture field decomposition. It has K readout ports and M input ports that can interface with an aperture antenna. The upper left block of the analog signal processor's S-matrix is $0^{K\textrm{x}K}$ indicating that its input ports are impedance matched and decoupled. The block $S_{io}^{K\textrm{x}M}$ performs the inner product of the aperture field with each of the aperture basis functions, and produces the weighting coefficients at each of the readout ports. The elements of $S_{oo}^{M\textrm{x}M}$ are free variables, and are neglected in the design.}
\label{fig-analog-signal-processor}
\end{figure}
\begin{figure}[!t]
\centerline{\includegraphics[width=\columnwidth]{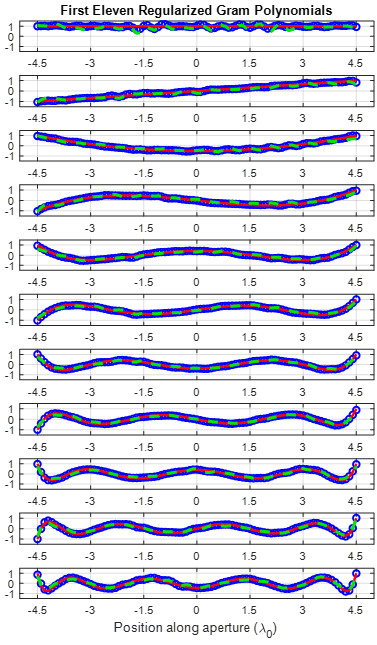}}
\caption{The eleven aperture basis functions (first eleven regularized Gram polynomials) used in the analog signal processor. The Gram polynomials have been normalized and plotted on a common scale. The solid red lines are the ideal Gram polynomials, the blue circles are the realized voltages across the aperture using the circuit network solver, and the green dashed line represent the voltages across the aperture in the full wave simulation.}
\label{GramPoly}
\end{figure}
\begin{figure}[!t]
\centerline{\includegraphics[width=\columnwidth]{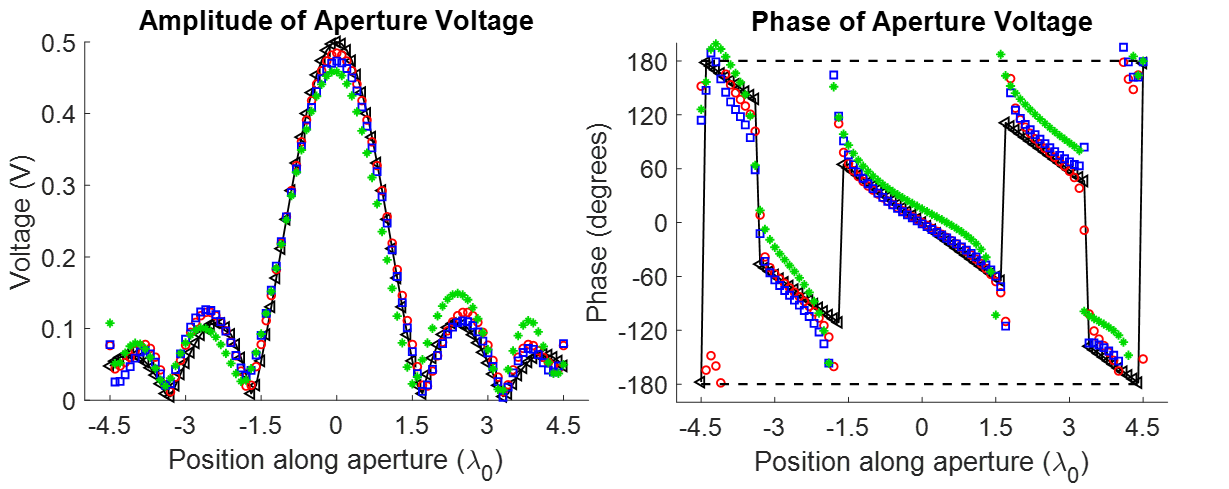}}
\caption{Comparison of the incident field profile to its approximation using the first eleven Gram polynomials with the idealized weighting coefficients and the weighting coefficients from the planar microstrip network (metastructure) computed using the circuit network solver and from the full-wave simulation. The incident aperture field is given by \eqref{Approx}. Black triangles: incident field, Red circles: reconstructed field using the idealized weighting coefficients, Blue squares: reconstructed field using the weighting coefficients from the circuit network solver, Green asterisks: reconstructed field using the weighting coefficients from the full-wave simulation.}
\label{Approx_field}
\end{figure}
\begin{figure}[!t]
\centerline{\includegraphics[width=\columnwidth]{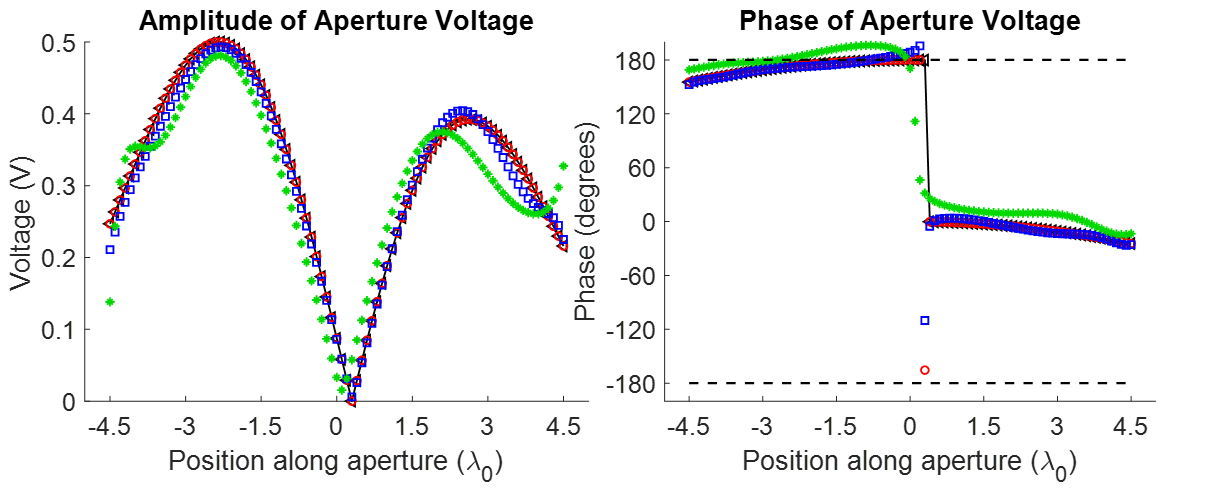}}
\caption{Comparison of the incident field profile to its approximation using the first eleven Gram polynomials with the idealized weighting coefficients and the weighting coefficients from the planar microstrip network (metastructure) using the circuit network solver and from the full-wave simulation. The incident aperture field is given by \eqref{Approx1}. Black triangles: incident field, Red circles: reconstructed field using the idealized weighting coefficients, Blue squares: reconstructed field using the weighting coefficients from the circuit network solver, Green asterisks: reconstructed field using the weighting coefficients from the full-wave simulation.}
\label{Approx_field1}
\end{figure}
\section{Analog Signal Processor}
In this section, an analog signal processor is designed to demonstrate the design method's ability to realize a variety of aperture fields. The analog signal processor samples an incident aperture field at its input ports and decomposes it into a set of aperture basis functions, and outputs the complex-valued weighting coefficient for each basis function, see Fig. \ref{fig-analog-signal-processor}. The coefficients are extracted from the network by observing the amplitude and phase of the voltage present at the readout port associated with each basis function. The extracted weighting coefficients can then be used to numerically reconstruct the aperture field using the known aperture basis functions.\\
\indent The aperture basis functions used in this example are a set of discrete orthogonal polynomials with equal Euclidean norm (ensures conservation of power). Specifically, the basis functions are a regularized version of the first eleven Gram or discrete Chebyshev polynomials. The Gram polynomials are selected since they are close approximations to the minimax polynomials used for approximating functions with finite sets of polynomials. The aperture basis functions are produced using the Discrete Orthogonal Polynomial toolbox in MATLAB \cite{DOP}, and are depicted in Fig. \ref{GramPoly}. Here, eleven basis functions are used, and it is demonstrated that eleven basis functions are sufficient to approximate some non-trivial aperture fields. However, if more accuracy is needed for highly oscillatory or aperture fields with discontinuities additional basis functions can be included by making the network larger or adjusting the spacings between the readout ports.
\subsection{Design Specifications}
The analog signal processor is implemented as a planar microstrip network. It can interface with an aperture antenna that operates at $10\textrm{GHz}$ and has a width of $W_{ap} = 9.1\lambda_0 = 27.3 \textrm{cm}$. The network has the same width as the antenna aperture and has a depth of $h=3\lambda_0=9 \textrm{cm}$. The microstrip unit cell shown in Fig. \ref{Unit_cell} is used to design the network. The unit cell size is $d=\lambda_0/10=3\textrm{mm}$, and the corresponding grid has the following dimensions, $N = 91$ and $M = 30$. Each of the eleven aperture basis functions are assigned to a $50 \Omega$ readout port. The readout ports are located along the input plane, shown in Fig. \ref{Ymatrix_network}, and are spaced by $0.7\lambda_0$, starting from the center line of the device. To ensure that each aperture basis function is associated with one readout port, all of the readout ports are required to be isolated from each other. The aperture field input ports, which interface with the aperture antenna, are located along the output plane, shown in Fig. \ref{Ymatrix_network}. Similar to the beamformer example, the aperture antenna's input ports have an input impedance of $140\Omega$ at broadside so, the aperture field input ports are matched to $140\Omega$ for each basis function. All of the remaining ports besides the readout and aperture field input ports are open-circuited during the design process.
\subsection{Optimization and Results}
\indent To design the network, the optimization routine is provided with the desired inputs (aperture basis functions) and readout voltages (single port excitations), the unit cell model, the terminations, and a initial set of 16,380 design variables. The design process took approximately 10 hours on a personal computer (i7-9700 CPU @ 3GHz w/8 cores with 64GB RAM) to produce a satisfactory design. The designed network's aperture basis functions are shown in Fig. \ref{GramPoly}. Excellent agreement between the desired and realized aperture basis functions is observed. The readouts are well matched with a maximum reflectance of $-19.2 \textrm{dB}$ for the seventh polynomial's readout port. The worst case isolation between the readout ports is $28.8 \textrm{dB}$, occurring between the readout port for the first polynomial and the readout port for the fifth polynomial.\\
\indent Two different incident aperture fields are used to test the analog signal processor's ability to decompose an incident field. The first aperture field is given by,
\begin{equation} \label{Approx}
V_{ap}(x) = 0.5\sinc(20x)e^{-jk_0\sin(\pi/6)x} \ \ V
\end{equation}
and the second incident aperture field is given by,
\begin{equation} \label{Approx1}
V_{ap}(x)=9.7\cos^5(\pi x/W_{ap})(x-0.009)e^{-j0.87k_0 x^2}
\end{equation}
where x is the position along the aperture in meters and $k_0$ is the free space wavenumber at 10GHz. The network is tested by solving for the complex weighting coefficients produced by the network for each incident aperture field using the circuit network solver. This is done by terminating the readout ports in $50 \Omega$ and the input ports in $140 \Omega$, and exciting the input ports with either \eqref{Approx} or \eqref{Approx1}. The reconstructed fields are calculated in MATLAB using the complex weighting coefficients produced by the network and the ideal aperture basis functions, and are shown in Fig. \ref{Approx_field} for \eqref{Approx} and Fig. \ref{Approx_field1} for \eqref{Approx1}. They are compared to the exact aperture fields as well, as idealized approximations of the aperture field. The weighting coefficients for the idealized approximation are computed by taking the inner product of the incident field with the first eleven Gram polynomials. The amplitude and phase of the reconstructed aperture fields show good agreement in both cases. The largest errors between the reconstructed and the exact amplitude and phase profiles occur near discontinuities or where the derivative changes sign. In these regions the reconstructed and idealized approximations of the aperture field show good agreement. Indicating that these errors are largely due to the number of basis functions used rather than issues with the design itself.\\
\indent For full-wave verification of the analog signal processor's performance, it is simulated in Keysight Momentum. The full-wave simulation took $\sim250$ hours to complete and the results for the aperture basis functions are shown in Fig. \ref{GramPoly}. Some variations in the aperture basis functions are observed but, overall the performance matches the circuit network solver quite well. The readouts are well matched with a maximum reflectance of $-12.7 \textrm{dB}$ for the fifth polynomial's readout port. The worst case isolation between the readout ports is $22 \textrm{dB}$, occurring between the readout port for the fourth polynomial and the readout port for the sixth polynomial. The full-wave solution for the device's scattering matrix is then used to reconstruct the aperture fields given by \eqref{Approx} and \eqref{Approx1} and the results are shown in Fig. \ref{Approx_field} and Fig. \ref{Approx_field1}, respectively. Again the results match quite well but, some errors are observed due to the errors in the aperture basis functions. The largest error is seen in the reconstructed amplitude of \eqref{Approx_field1} around the position $2\lambda_0$. This is largely due to errors present in the full-wave results for the fourth aperture basis function around this position since, \eqref{Approx1} has a significant component along this basis function.
\section{Conclusion}
In this paper, an inverse-design procedure for multi-input multi-output (MIMO) metastructured devices was provided. The design procedure uses a fast 2-D circuit network solver in conjunction with a gradient-based optimization routine to produce devices with desired MIMO functions. Since the gradient must be calculated at every step of the optimization routine, and metastructures have a large number of design variables, the adjoint variable method is used to calculate the gradient. The computational efficiency gained by using the fast 2-D circuit-based solver and the adjoint variable method enables the design procedure to realize electrically large aperiodic MIMO metastructures.\\
\indent The efficacy of the design procedure was then demonstrated through the design of a planar antenna beamformer and an analog signal processor for aperture field decomposition. The beamformer supports the simultaneous excitation of nine beams, contains approximately 6000 design variables, and took approximately six hours to design. The analog signal processor uses eleven orthogonal aperture basis functions to decompose incident field profiles. It contains approximately 16,400 design variables and took approximately 10 hours to design. Both of the devices were implemented in microstrip technology and their performances were verified using the Keysight method of moments solver Momentum.\\

\begin{IEEEbiography}[{\includegraphics[width=1in,height=1.25in,clip,keepaspectratio]{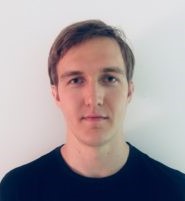}}]{Luke Szymanski} (Graduate Student Member, IEEE) received the B. Sc. in electrical engineering from the University of Texas at Dallas in 2015 and the M. Sc. in electrical and computer engineering from the University of Michigan, Ann Arbor in 2020. In 2016, he joined Professor Grbic's group at the University of Michigan as a Ph.D. student. His research interests include analytical electromagnetics and optimization-based inverse design of metamaterial devices and metasurfaces.
\end{IEEEbiography}

\begin{IEEEbiography}[{\includegraphics[width=1in,height=1.25in,clip,keepaspectratio]{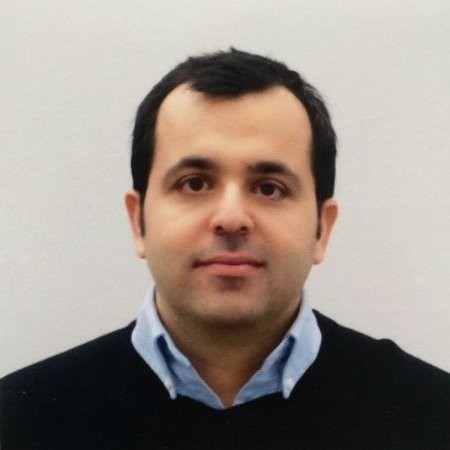}}]{Gurkan Gok} (Member,IEEE) received the B.S.E. degree in Electrical Engineering from Middle East Technical University in 2008 and the M.S.E. and Ph.D. degrees in electrical engineering from the University of Michigan, Ann Arbor, MI, USA, in 2010 and 2014, respectively. He is currently a staff member at the Raytheon Research Center, East Hartford, CT. His research includes antennas, microwave circuits, sensors and networks, additive manufacturing, metamaterial structures.

\end{IEEEbiography}

\begin{IEEEbiography}[{\includegraphics[width=1in,height=1.25in,clip,keepaspectratio]{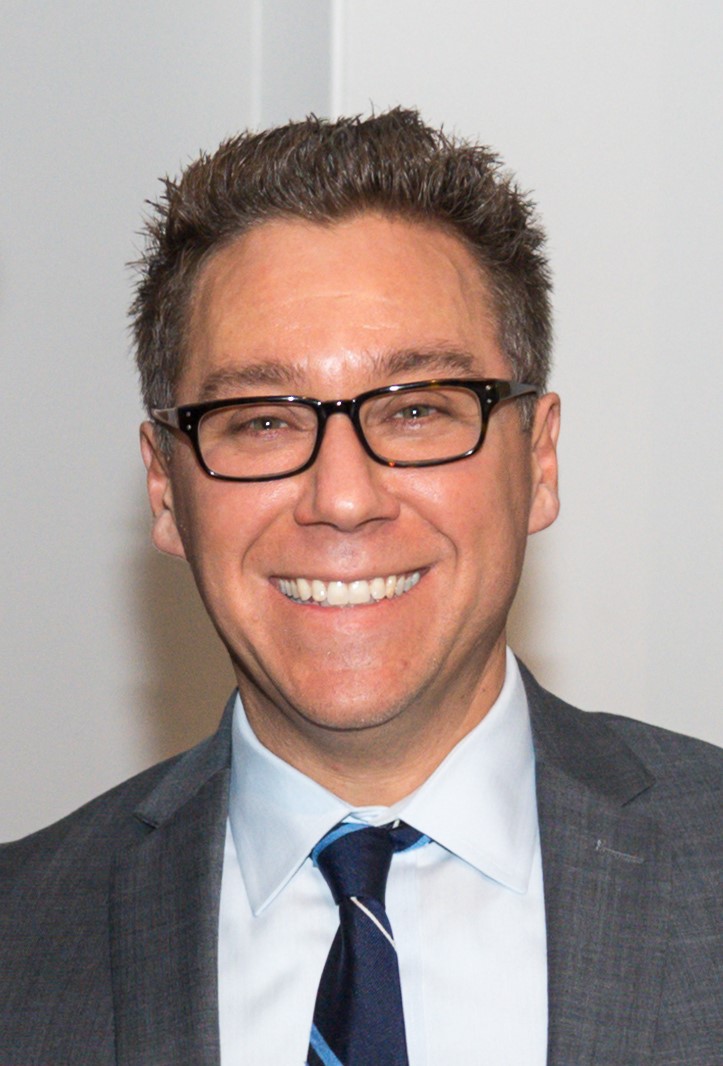}}]{Anthony Grbic} (Fellow, IEEE) received the B.A.Sc., M.A.Sc., and Ph.D. degrees in electrical engineering from the University of Toronto, Toronto, ON, Canada, in 1998, 2000, and
2005, respectively. In January 2006, he joined the Department of Electrical Engineering and Computer Science, University of Michigan, Ann Arbor, MI, USA, where he is currently a Professor. His research
interests include engineered electromagnetic structures (metamaterials, metasurfaces, electromagnetic band-gap materials, frequency-selective surfaces), plasmonics, antennas, microwave circuits, wireless power transmission, and analytical electromagnetics/optics. Dr. Grbic served as Technical Program Co-Chair in 2012 and Topic Co-Chair in 2016 and 2017 for the IEEE International Symposium on Antennas and Propagation and USNC-URSI National Radio Science Meeting. He was an Associate Editor for IEEE Antennas and Wireless Propagation Letters from 2010 to 2015. He is currently the Vice Chair of AP-S Technical Activities, Trident Chapter, IEEE Southeastern Michigan Section. Dr. Grbic was the recipient of AFOSR Young Investigator Award as well as NSF Faculty Early Career Development Award in 2008, the Presidential Early Career Award for Scientists and Engineers in January 2010. He also received an Outstanding Young Engineer Award from the IEEE Microwave Theory and Techniques Society, a Henry Russel Award from the University of Michigan, and a Booker Fellowship from the United States National Committee of the International Union of Radio Science in 2011. He was the inaugural recipient of the Ernest and Bettine Kuh Distinguished Faculty Scholar Award in the Department of
Electrical and Computer Science, University of Michigan in 2012.
\end{IEEEbiography}


\begin{thebibliography}{00}

\bibitem{Ozcan}
A. Ozcan, et. al. ``All-optical machine learning using diffractive deep neural networks'' \emph{Science}, 361, 1004–1008 (2018)

\bibitem{Backer}
A. Backer, ``Computational inverse design for cascaded systems of metasurface optics,'' \emph{Optics Express}, Vol. 27, Issue 21,  pp. 30308-30331 (2019)

\bibitem{Engheta}
Estakhri, N. M., Edwards, B., Engheta, N. ``Inverse-designed metastructures that solve equations.'' \emph{Science}, 363, 1333–1338 (2019)

\bibitem{Tierney}
B. B. Tierney and A. Grbic, ``Designing Anisotropic Inhomogeneous Metamaterial Devices Through Optimization,'' \emph{Antennas and Propagation IEEE Transactions on}, vol. 67, no. 2, pp. 998-1009, 2019.

\bibitem{Sanford}
J. R. Sanford, ``Design of a Miniature Reactive Beam Forming Network,'' \emph{2019 IEEE International Symposium on Antennas and Propagation and USNC-URSI Radio Science Meeting}, Atlanta, GA, USA, 2019, pp. 1357-1358

\bibitem{Johnson}
S. G. Johnson, et. al. ``Inverse design of large-area metasurfaces,'' \emph{Optics Express},  Vol. 26, Issue 26, pp. 33732-33747 (2018)

\bibitem{Yablonovitch}
E. Yablonovitch et. al., ``Adjoint shape optimization applied to electromagnetic design,'' \emph{Opt. Express} 21, pp. 21693-21701 (2013)

\bibitem{Siddiqui}
O.F. Siddiqui, ``The Forward Transmission Matrix (FTM) Method for S-Parameter Analysis of Microwave Circuits and Their Metamaterial Counterparts,'' \emph{Progress In Electromagnetics Research B}, Vol. 66, 123-141, 2016. doi:10.2528/PIERB16012101

\bibitem{Mesa}
F. Mesa, G. Valerio, R. Rodriguez-Berral and O. Quevedo-Teruel, ``Simulation-Assisted Efficient Computation of the Dispersion Diagram of Periodic Structures: A Comprehensive Overview With Applications to Filters, Leaky-Wave Antennas and Metasurfaces,'' in \emph{IEEE Antennas and Propagation Magazine}, doi: 10.1109/MAP.2020.3003210.

\bibitem{Alsolamy}
F. Alsolamy and A. Grbic, ``Modal Network Formulation for the Analysis and Design of Mode-Converting Metasurfaces in Cylindrical Waveguides,'' in \emph{IEEE Trans. on Antennas Propag.,} doi: 10.1109/TAP.2020.3048590.

\bibitem{Johnson_adjoint}
S. G. Johnson, ``Notes on Adjoint Methods for 18.335'' https://math.mit.edu/$\sim$stevenj/18.336/adjoint.pdf (Last accessed Dec. 16, 2020).

\bibitem{Luneburg}
R.K. Luneburg, ``The Mathematical Theory of Optics,'' Brown University Press, Providence, R. I., pp. 208-213; 1944.

\bibitem{Rotman}
W. Rotman and R. F. Turner, ``Wide-angle microwave lens for line
source applications,'' \emph{IEEE Trans. Antennas Propag.}, vol. AP-11, no.
11, pp. 623–632, Nov. 1963.

\bibitem{Gok}
G. Gok and A. Grbic, ``A printed antenna beam former implemented using tensor transmission-line metamaterials,'' \emph{in Proc. IEEE Antennas Propag. Soc. Int. Symp.}, Jul. 2014, pp. 765–766

\bibitem{Butler}
J. Butler and R. Lowe, ``Beam forming matrix simplifies design of electronically scanned antennas,'' \emph{Electron. Design}, vol. 9, pp. 170-173, Apr. 1961.

\bibitem{Hansen}
RC Hansen, ``Linear connected arrays [coupled dipole arrays],'' in \emph{IEEE Antennas and Wireless Propagation Letters}, vol. 3, pp. 154-156, 2004, doi: 10.1109/LAWP.2004.832125.

\bibitem{Allen} J. L. Allen, ``A Theoretical Limitation on the Formation of Lossless Multiple Beams in Linear Arrays,'' in \emph{IRE Trans. Antennas Propag.,} vol. 9, no. 4, pp. 350--352. July, 1961.

\bibitem{White} W. D. White, ``Pattern Limitations in Multiple-Beam Antennas,'' in \emph{IRE Trans. Antennas Propag.,} vol. 10, no. 4, pp. 430--436. July, 1962

\bibitem{Stein} S. Stein, ``On Cross Coupling in Multiple-Beam Antennas,'' in \emph{IRE Trans. Antennas Propag.,} vol. 10, no. 5,  pp. 548--557. Sept., 1962

\bibitem{Pestourie}
Pestourie, R., et al. ``Active learning of deep surrogates for PDEs: application to metasurface design.'' \emph{npj Comput Mater} 6, 164 (2020).

\bibitem{Holloway}
C. L. Holloway, et al., ``Characterizing Metasurfaces/Metafilms: The Connection Between Surface Susceptibilities and Effective Material Properties,'' in \emph{IEEE Antennas and Wireless Propagation Letters}, vol. 10, pp. 1507-1511, 2011

\bibitem{Pozar}
D.M. Pozar, \emph{Microwave Engineering}, 4th ed. New York:Wiley, 2011 pp. 422--426.

\bibitem{DOP}
Matthew Harker, Paul O'Leary, (2020). Discrete Orthogonal Polynomial Toolbox: DOPBox Version 1.8 (https://www.mathworks.com/matlabcentral/fileexchange/41250-discrete-orthogonal-polynomial-toolbox-dopbox-version-1-8), MATLAB Central File Exchange. Retrieved December 21, 2020.
\end{thebibliography}
\end{document}